\documentclass[twocolumn,aps,showpacs,superscriptaddress,nofootinbib]{revtex4-1}  

\usepackage[T1]{fontenc}
\usepackage[utf8]{inputenc}
\usepackage{lmodern}
\usepackage{amsmath}
\usepackage{amsfonts}
\usepackage{booktabs}
\usepackage{tabularx}
\usepackage{graphicx}
\usepackage{todonotes}
\usepackage[english]{babel}
\usepackage{hyperref}
\usepackage[caption=false]{subfig}
\usepackage{natbib}

\begin{document}

\title{
Gravitational Higgs Mechanism in Neutron Star Interiors}

\author{Andrew Coates}
\affiliation{School of Mathematical Sciences, University of Nottingham, University Park, Nottingham, NG7 2RD, UK}
\author{Michael W.~Horbatsch}
\affiliation{School of Mathematical Sciences, University of Nottingham, University Park, Nottingham, NG7 2RD, UK}
\author{Thomas P.~Sotiriou}
\affiliation{School of Mathematical Sciences, University of Nottingham, University Park, Nottingham, NG7 2RD, UK}
\affiliation{School of Physics and Astronomy, University of Nottingham, University Park, Nottingham, NG7 2RD, UK} 

\date{\today}

\begin{abstract}
We suggest that nonminimally coupled scalar fields can lead to modifications of the microphysics in the interiors of relativistic stars. As a concrete example, 
we consider the generation of a non-zero photon mass in such high-density environments. 
This is achieved by means of a light gravitational scalar,
and the scalarization phase transition in 
scalar-tensor theories of gravitation. Two distinct models are presented, 
and phenomenological implications are briefly discussed.
\end{abstract}

\maketitle

Scalar-tensor theories \cite{Faraoni:2004pi,Fujii:2003pa,Damour:1992we,Chiba:1997ms} can be thought of as theories of gravity with an additional scalar field $\Phi$ that couples non minimally to the metric $\tilde{g}_{\mu\nu}$ but does not couple to the matter fields, $\Psi^A$. The latter couple minimally to $\tilde{g}_{\mu\nu}$ only. In this representation, known as the Jordan frame, the action reads 
\begin{eqnarray}
S[\tilde{g}_{\mu \nu} , \Phi] &=& \frac{1}{16\pi}\int \mathrm{d}^{4}x \sqrt{-\tilde{g}} 
\left(
\Phi \tilde{R} - \frac{\omega(\Phi)}{\Phi}\tilde{g}^{\mu \nu} \partial_{\mu}\Phi
\partial_{\nu}\Phi 
\right) 
\nonumber
\\
&&+ S_{\rm m}[\Psi^{A}, \tilde{g}_{\mu \nu}]\,,
\end{eqnarray}
and the weak equivalence principle (WEP) is manifest. Here $\Phi$ has the interpretation of 
a varying inverse gravitational constant,  
$S_{\rm m}$ denotes the matter action, and $\tilde{R}$ is the Ricci scalar of $\tilde{g}_{\mu\nu}$. 

One can also reformulate this action in terms of another metric and a redefined scalar field, in the so called Einstein frame. The scalar field  and the metric in this frame are related to their Jordan frame counterparts by,
\begin{equation}
\Phi = [G_{\star}A^{2}(\phi)]^{-1}, \qquad \tilde{g}_{\mu\nu}=A^2(\phi)g_{\mu\nu}\,,
\end{equation}
where $G_{\star}$ is a bare gravitational constant. 
The form of \(A^2(\phi)\) is determined by the choice of \(\omega(\Phi)\) and the requirement that the kinetic term for \(\phi\) be canonical.
That is, the Einstein frame action reads
\begin{align}
S=&\frac{1}{16\pi G_*}\int\mathrm{d}^4x\sqrt{-g}\left[R-2g^{\mu\nu}\left(\partial_\mu\phi\right)\left(\partial_\nu\phi\right)\right]\nonumber\\
&+S_{\rm m}[\Psi^A,A^2(\phi)g_{\mu\nu}],
\end{align}
where \(R\) is the Ricci scalar of \(g_{\mu\nu}\). In the absence of matter the theory clearly reduces to general relativity (GR) with a minimally coupled scalar field. In this representation the deviation from GR is encoded in the nonmiminal coupling between the matter and $\phi$.

After some manipulations the field equation for the scalar field can be put into the form \cite{Faraoni:2004pi,Fujii:2003pa,Damour:1992we,Chiba:1997ms,Damour:1993hw}\begin{equation}
\label{seq}
\Box \phi+4\pi G_*T \frac{d}{d\phi} \log A(\phi)=0\,,
\end{equation}
where $T$ is the trace of the Einstein frame stress-energy tensor
\begin{equation}
T_{\mu \nu} =
- \frac{2}{\sqrt{-g}}
\frac{\delta S_{\rm m}}{\delta g^{\mu\nu}}\,.
\end{equation}
Theories in which $dA(\phi_0)/d\phi=0$ for some constant $\phi_0$, admit GR solutions with a trivial scalar configuration, as the scalar's equation is trivially satisfied and $g_{\mu\nu}$ effectively satisfies Einstein's equation with a rescaled gravitational coupling. Note that such theories have $\omega(\Phi_0)\to \infty$ in the Jordan frame, where $\Phi_0\equiv \Phi(\phi_0)$ (see Ref.~\cite{Sotiriou:2015lxa} for a more detailed discussion).

Certain theories in this class exhibit a remarkable property dubbed spontaneous scalarization \cite{Damour:1993hw,Damour:1996ke,Chiba:1997ms}. It is convenient to expand  
the logarithmic derivative of the conformal factor around $\phi=\phi_0$ as
\begin{equation}
\label{expansion}
\log A(\phi)=A_0+\beta_0(\phi-\phi_0)^2/2+\cdots.
\end{equation}
In and around stars of relatively low densities, such as the Sun, the scalar remains at the trivial configuration, $\phi=\phi_0$, and the metric is that of GR. As a result the theory is indistinguishable from GR in the weak field limit. However, for \(\beta_0\lesssim-4\), compact stars above a threshold central density undergo a phase transition and develop a large scalar charge, even in the absence of an external scalar environment \cite{Damour:1993hw}. This behaviour is of particular interest
as it underscores the importance of constraining deviation from GR in the strong field regime. 

At the perturbative level spontaneous scalarization can be seen as a tachyonic instability \cite{Damour:1993hw}. The coupling between the scalar and $T$ in eq.~(\ref{seq}) generates a negative mass for scalar perturbations around $\phi=\phi_0$. The end point of this instability is the scalarized solution, which exhibits no such instability. This perturbative manifestation allows one to determine with good certainty whether spontaneous scalarization occurs without performing a more complete nonperturbative analysis. 

Spontaneous scalarization changes the structure of compact stars \cite{Damour:1993hw,Damour:1996ke,Chiba:1997ms} and, as a result, it has recently been severely constrained by binary pulsar observations \cite{Damour:1998jk,Freire:2012mg}. However, it does so without actually affecting the microphysics in the star. This is manifest in the Jordan frame picture, where the scalar field $\Phi$ does not couple directly to matter. Hence, when it develops a nontrivial profile it does act as a source for the gravitational field and it changes the binding energy of the star, but it does not change the properties of the matter fields. In particular, matter  can still be described effectively as a fluid with a certain  equation of state (EOS) and this EOS can be determined without any reference to $\Phi$.

In this paper we wish to consider the more intriguing possibility that the existence of a scalar field can actually change the behaviour of matter inside a neutron star. This can be achieved without violating WEP constraints if the scalar is coupled to matter in a way such that: (i) this coupling vanishes to the desired order in perturbation theory around unscalarized solutions; (ii) it comes to life once scalarization has occurred. This coupling could then substantially modify the masses and/or coupling constants of standard model fields in scalarized environments, such as neutron star interiors. For concreteness and as a proof of principle, we will consider the case of the electromagnetic field and we will present two concrete models for photon mass generation 
and amplification. In both models, the mechanism underlying the variation of fundamental 
constants is the scalarization phase transition in 
scalar-tensor theories of gravity \cite{Damour:1993hw}.\footnote{
The possibility that these masses and couplings are not  constant, but 
rather vary throughout space-time has been extensively explored in a different context.
For reviews, see Refs.~\cite{Uzan:2002vq,19611,19629,19686,18935,19813,19917,Goldhaber:2008xy,lrr-2011-2}. 
}

Since the toy models considered in this work deal with 
photon mass generation and amplification, the starting point will be the action of the electromagnetic field,
\begin{equation}
S_{\rm EM}[A_{\mu},\tilde{g}_{\mu \nu}] = 
S_{\rm EM}[A_{\mu},g_{\mu \nu}] = 
- \frac{1}{4} \int d^{4} x \sqrt{-g} F_{\mu \nu} F^{\mu \nu} \,,
\end{equation}
where 
$F_{\mu \nu} = \partial_{\mu} A_{\nu} - \partial_{\nu}A_{\mu}$ 
is the usual field strength tensor. This action is invariant
under the 
$U(1)$ gauge symmetry $A_{\mu} \to A_{\mu} + \partial_{\mu}\lambda$. 
Following the motivation outlined so far, the simplest model to consider would appear to be one where we introduce an additional Proca-like coupling between the scalar field and the photon. For instance, consider the case where the unscalarized solution corresponds to \(\phi=\phi_0=0\) and one adds to the matter action the term,
\begin{equation}
\label{massterm}
\frac{1}{2}\left(m\phi\right)^{2}A_\mu A^\mu.
\end{equation}
Note that setting $\phi_0=0$ amounts to a global shift of the scalar, so it can be done without loss of generality.
The term in eq.~\eqref{massterm} clearly generates a mass for electromagnetic perturbations in a scalarized setting and having an even power of \(\phi\) guarantees that the mass of the photon will not be negative. At the same time, it does not introduce any modification at linear order (always counting at the level of the field equations) in the unscalarized case as,
\begin{equation}
\frac{1}{2}\left(m\phi\right)^{2}A_\mu A^\mu\propto (\delta\phi)^2 \delta A^\mu \delta A_\mu,
\end{equation} 
around \(\phi=0=A^\mu\). However, this model, despite being potentially interesting, is actually not perturbative around \(\phi=0\). This is unappealing because it casts doubt on whether the scalarized and the unscalarized phases are continuously connected. 

To understand this issue better we should first review the case of the standard Proca field, with Lagrangian,
\begin{equation}
\mathcal{L}=-\frac{1}{4}F_{\mu\nu}F^{\mu\nu}-\frac{1}{2}m^2A_\mu A^\mu.
\end{equation}
When one considers the polarizations of the vector field one sees that the logitudinal mode disappears in the limit \(m\to 0\) and thus one expects a discontinuity. To reinstate the \(U(1)\) gauge symmetry and investigate this limit more carefully one can introduce the Stueckelberg field \(\psi\) \cite{Ruegg:2003ps}. If under a \(U(1)\) transformation
\begin{align}
A_\mu \to A_\mu +\partial_\mu \lambda, \qquad  \psi \to \psi - m \lambda,
\end{align}
then the Lagrangian,
\begin{equation}
\mathcal{L}_S=-\frac{1}{4}F_{\mu\nu}F^{\mu\nu}-\frac{1}{2}\left(m A_\mu +\partial_\mu \psi\right)^2,
\end{equation}
is gauge invariant. Now when one takes \(m\to 0\) the ``longitudinal mode", i.e. \(\psi\), decouples from the theory and thus no true discontinuity exists. Indeed the original issue can be interpreted as the choice \(\psi=0\) being a bad gauge for addressing this question.

To generalize this to the case of interest there appear to be two possibilities for introducing a Stueckelberg field. The first option is to keep the original gauge transformation, in which case the mass-like term becomes,
\begin{equation}
\label{m1}
\frac{\phi^2}{2}\left(m A_\mu +\partial_\mu \psi\right)^2.
\end{equation}
Despite \(m\to 0\) still being a proper decoupling limit, if one attempts to perturb around \(\phi=0\) the kinetic term for \(\psi\) does not appear until cubic order in perturbation theory. This discontinuity in the degrees of freedom between different orders of perturbation theory seems to challenge the validity of such a treatment.

The second option would be to include \(\phi\) in the gauge transformation, that is, to attempt to follow the Stueckelberg perscription but treating \(m\phi\) rather than \(m\) as the ``mass parameter". This means replacing the transformation \(\psi\to \psi - m\lambda\) by,
\begin{equation}
\psi \to \psi -m\phi\lambda.
\end{equation}
In this case there are additional counter-terms and the mass-like term is,
\begin{equation}\label{phidepgauge}
\left(\partial_\mu \psi - \psi\partial_\mu\log\phi+m\phi A_\mu\right)^2.
\end{equation}
Due to the term containing $\log\phi$  perturbative treatment around \(\phi=0\) is compromised, even though \(\phi=0\) is still a perfectly acceptable asymptotic value for $1/r^n$,
exponential and Yukawa decays.
It is worth pointing out that this prescription for introducing the Stueckelberg field is  actually related to the previous via the field redefinition  \(\chi=\psi/\phi\) that transforms eq.~\eqref{phidepgauge}  into eq.~\eqref{m1} with $\psi$ replaced by $\chi$.

A straightforward (though perhaps not unique) way to circumvent the discontinuity issue in this model is the following. Instead of generating a mass, one can start with a mass that is tuned to be undetectably small in unscalarized backgrounds and simply enhance it around scalarized backgrounds.
Consider the  Einstein frame Lagrangian
\begin{align}\label{enhancedmasslag}
\frac{\mathcal{L}_S}{\sqrt{-g}} = &\frac{1}{4\pi G_*}\left( \frac{R}{4} -\frac{1}{2}\left(\partial_\mu\phi\right)^2\right) -\frac{1}{4}F_{\mu\nu}F^{\mu\nu}\nonumber\\
&-\frac{A^{2}(\phi)}{2}(1+f(\phi))\left(m A_\mu +\partial_\mu \psi\right)^2
\end{align}
where  \(f(0)=0\) and $f$ is positive for all other arguments. In the Jordan frame for the (rest of the) matter, for invertible \(A\), the Lagrangian takes the form,
\begin{align}
\frac{\mathcal{L}_S}{\sqrt{-\tilde{g}}} =& \frac{1}{16\pi}\left(\Phi \tilde{R} -\frac{\omega(\Phi)}{\Phi}\left(\partial_\mu \Phi\right)^2\right) -\frac{1}{4}F_{\mu\nu}F^{\mu\nu}\nonumber\\
&-\frac{1}{2}(1+h(\Phi))\left(m A_\mu +\partial_\mu \psi\right)^2.
\end{align}
where, \(\Phi = G_*^{-1}A^{-2}(\phi)\), \(h(\Phi)=f\left(\phi(\Phi)\right)\). In other words, as one may expect, the Jordan frame photon mass (squared) is \(m^2(1+f(\phi))\). Working perturbatively around $\phi=0$ (the unscalarized solution) it is clear that the mass of the photon will be determined by the value of $m$ and so it can be tuned to a desired value that would avoid any know constraint. On the other hand, when $\phi$ has a nontrivial configuration the effective mass of the photon is clearly modified. Enhancing it to the desired value is just a matter of making a suitable choice of $f$. 

What remains is to argue that the presence of the mass term will not prevent scalarization from occurring. In order to study this process in its full glory and determine the stellar structure, it is necessary to numerically integrate the equations of motion derived from the Lagrangian in eq.~\eqref{enhancedmasslag}. However, with some approximations, it is possible to obtain an analytical understanding of the scalarization process. In particular, one can look for the standard sign for spontaneous scalarization at the perturbative level, which is the onset of a tachyonic instability for $\phi$ once the star reaches a threshold compactness.

Indeed, so long as $f$ is chosen to vanish at least to quadratic order in perturbation theory, the original perturbative calculation of Ref.~\cite{Damour:1993hw} applies here. For concreteness and without significant loss of generality, we shall take \(A^2(\phi)=\exp(\beta \phi^2)\). Recall that if one expands a more general choice for \(A^2(\phi)\) as in eq.~\eqref{expansion}, it is the value of \(\beta_0\) that controls how effective scalarization is. Consider the scalar's field equation, eq.~\eqref{seq}, sourced by a constant (in the Jordan frame) matter density, \(\rho\), within some radius, \(R\), and take the metric to be that of flat space in order to decouple the tensor field equations 
from the scalar equation. In terms of the dimensionless compactness parameter,
\(s=G_*M/R\) (\(\sim 0.2\) for neutron stars) and defining \(u=r/R\) the scalar's equation of motion becomes,
\begin{equation}
\frac{\mathrm{d}^2\phi}{\mathrm{d}u^2}+\frac{2}{u}\frac{\mathrm{d}\phi}{\mathrm{d}u}=3s\beta e^{2\beta\phi^2}\phi H(1-u).
\end{equation}
where \(H(x)\) is the Heaviside step function. Solving this equation in a small-amplitude expansion for \(\phi\) to sub-leading order, and matching \(\phi\) and \(\phi'\) at the stellar boundary one finds, for \(\beta<0\), a non-trivial scalar profile \(\phi(u)=A\, \mathrm{sinc}(\tau u)+\mathcal{O}(A^3)\) with \(\tau=\sqrt{3s|\beta|}\) and amplitude \(A\sim (s-s_*)^{1/2}\), whenever the compactness, $s$, exceeds the scalarization 
threshold \(s_*=\pi^2/(12|\beta|)\). The mass term will not contribute to $\phi$'s equation due to our assumption that $f$ vanishes to quadratic order in perturbations. Hence it cannot quench this instability. 

As a second example, we now turn to another model, which will  allow for an actual generation of a mass and can be straightforwardly generalized
to other gauge fields. Since scalarization appears, perturbatively, as a tachyonic instability of the \(\phi=0\) configuration around compact objects, it is quite tempting to consider a Higgs mechanism where the coupling to matter (or non-minimal coupling in the Jordan frame) replaces the Higgs potential in its role.
Consider a scalar-tensor theory with a complex charged scalar field 
$\phi$, and action 
\begin{eqnarray}
S &=& \frac{1}{4\pi G_{\star}}
\int d^4 x \sqrt{-g}
\left\{
\frac{R}{4}
- \frac{1}{2} g^{\mu \nu}
\overline{D_{\mu} \phi} D_{\nu} \phi \right\}
\\
&& - \frac{1}{4} \int d^4 x \sqrt{-g} F_{\mu \nu}F^{\mu \nu}
+ S_{\rm m}[A^{2}(\bar{\phi}\phi)g_{\mu \nu} ;\, A_{\mu} ;\, \Psi^{A} ] \,.\nonumber
\end{eqnarray}
Here $D_{\mu} \phi = \partial_{\mu} \phi - ieA_{\mu}\phi$ is the gauge
covariant derivative of the scalar, where 
the constant $e$ determines the coupling between the gravitational
scalar and the photon.

The conformal factor is taken to depend only on the modulus of the scalar, 
and is taken to be 
$A(\bar{\phi}\phi) = \exp (\tfrac{1}{2} \beta \bar{\phi}\phi)$, 
where $\beta$ is a constant. 
With these choices, the action is invariant under the
$U(1)$ gauge transformation
\begin{equation}
\phi \to \phi e^{ie\lambda} \,,
\qquad
A_{\mu} \to A_{\mu} + \partial_{\mu}\lambda \,.
\end{equation}
If this symmetry is spontaneously broken, and the scalar $\phi$
develops a non-zero vacuum expectation value, 
then the photon attains a mass of
\begin{equation}
m_{\gamma}^{2}(\bar{\phi}\phi) = \frac{e^2}{4\pi G_{\star}}
\bar{\phi}\phi \,. 
\end{equation}

Since this model allows for mass generation, let us study it in a bit more detail. The field equations are 
\begin{eqnarray}
&&( \Box  - e^2A_{\mu}A^{\mu}  - 2ieA^{\mu}\partial_{\mu}
\nonumber\\
\label{mod2_fieldeq_scalar}
&&\qquad\qquad-ie\nabla_{\mu}A^{\mu} ) \phi=- 4 \pi G_{\star}T \beta \phi  \,,\\
\label{Maxwell}
&&\nabla^{\mu}F_{\mu \nu} = 
J_{\nu} + J_{\nu}^{(\phi)}
+ m_{\gamma}^{2}(\bar{\phi}\phi) A_{\nu} \,,\\
&&G_{\mu \nu}
= 8 \pi G_{\star} 
\left(
T_{\mu \nu}
+ T_{\mu \nu}^{(\phi)}
+ T_{\mu\nu}^{(A)}
+ T_{\mu\nu}^{(\phi A)}
\right) \,,
\end{eqnarray}
where
\begin{eqnarray}
J_{\mu} &=& -\frac{1}{\sqrt{-g}}
\frac{\delta S_{\rm m}}{\delta A^{\mu}}\,,\\
T_{\mu \nu}^{(\phi)}&=& 
\frac{1}{4 \pi G_{\star}}
\left(
\partial_{( \mu}\bar{\phi} \partial_{\nu )}\phi
- \frac{1}{2} g_{\mu \nu}
g^{\lambda \sigma}
\partial_{\lambda} \bar{\phi}
\partial_{\sigma} \phi
\right) \,,\\
J_{\mu}^{(\phi)} &=&  \frac{ie}{8\pi G_{\star}} 
\left(
\bar{\phi} \partial_{\mu} \phi 
- \phi \partial_{\mu}\bar{\phi}
\right) \,,\\
T_{\mu \nu}^{(A)} &=& F_{\mu \lambda}F_{\nu}^{\hphantom{\nu}\lambda}
- \frac{1}{4}
g_{\mu \nu}
F_{\lambda \sigma} F^{\lambda \sigma}
\nonumber
\\
&&
+m_{\gamma}^{2}(\bar{\phi}\phi) 
\left(
A_{\mu}A_{\nu}
- \frac{1}{2}g_{\mu \nu}g^{\lambda \sigma}
A_{\lambda}A_{\sigma}
\right) \,,\\
T_{\mu \nu}^{(\phi A)} &=& 
2 \left(
J_{(\mu}^{(\phi)}A_{\nu )} 
- \frac{1}{2}g_{\mu \nu}g^{\lambda \sigma}
 J_{\lambda}^{(\phi)}A_{\sigma}
\right) \,.
\end{eqnarray}

In a static spherically-symmetric space-time, the line element has the form
\begin{equation}
g_{\mu \nu}dx^{\mu} dx^{\nu} = 
-f(r)dt^2 + h(r) dr^2 + k(r)d\Omega^2 \,,
\end{equation}
and the electromagnetic potential has the form
\begin{equation}
A^{\mu} = \left( A^{t}(r) \, , \, A^{r}(r) \, , \, 0 \, , \, 0 \right)  \,,
\end{equation}
and $\phi$ is purely radial.
In this geometry, the field equations \eqref{mod2_fieldeq_scalar}-\eqref{Maxwell} reduce to
\begin{eqnarray}
\label{ss1}
&&\phi''
+ {\cal A} \phi'
-{\cal B} h \phi 
= -4 \pi G_{\star} \beta  T h  \phi \,,\\
\label{ss2}
&&\left( \frac{f'}{f}A^{t} + (A^{t})'\right) '
+ \left( \frac{f'}{2f} - \frac{h'}{2h} + \frac{k'}{k} \right)
\left( \frac{f'}{f}A^{t} + (A^{t})'\right)
= \nonumber \\&&\qquad\qquad\qquad =- h J^{t} +  m_{\gamma}^{2}(\bar{\phi}\phi) h A^{t} \,,\\
\label{ss3}
&&J^{r} = m_{\gamma}^{2}(\bar{\phi}\phi)A^{r} + \frac{J_{r}^{(\phi)}}{h}  \,,
\qquad
J^{\theta} = J^{\phi} = 0 \,,
\end{eqnarray}
where
\begin{eqnarray}
&&{\cal A}\equiv \frac{f'}{2f} - \frac{h'}{2h} + \frac{k'}{k} - 2iehA^{r}\,,\\
&& {\cal B}\equiv  ie (A^{r})' + ie \left( \frac{h'}{2h} + \frac{f'}{2f} + \frac{k'}{k}\right) A^{r} \nonumber\\
&&\qquad\qquad + e^2 h (A^{r})^2 - e^2 f (A^{t})^2 \,.
\end{eqnarray}

If the matter described by $S_{\rm m}$ is electrically neutral, then the external current $J^{\mu}$ vanishes, and it follows from the 
above equations that 
\begin{equation}
\Im \left( \frac{\phi'}{\phi} \right) = e h A^{r} \,.
\end{equation}
In other words, the phase of $\phi$
is directly related to the radial component of the electromagnetic potential, 
and thus $A^{\mu}$ has only one dynamical degree of freedom. 
This is the concrete manifestation of gauge invariance in spherical symmetry.

Given that the scalar current source $J_{\mu}^{(\phi)}$ appears on the right-hand side of 
eq.~\eqref{Maxwell}, one might be tempted to conclude that scalarized stars 
necessarily have an electric charge. This would be highly problematic from a 
phenomenological point of view. However, it follows from the eqs.~\eqref{ss1}-\eqref{ss3} that, 
 in the static and 
spherically-symmetric case, the only non-trivial component of this current
can be absorbed by a gauge transformation. Hence, it is in principle 
possible to have scalarized stars with vanishing electric charge, but 
still a non-zero photon mass in their interior, allowing for interesting 
phenomenology.

 We will now demonstrate that the electric charge of scalarized stars is actually forced to vanish at the perturbative level.
We will resort to the same approximations and definitions as in the previous model. We will additionally define \(W=\sqrt{4\pi G_*}A^t\) and \(\epsilon=e^2R^2/(4\pi G_*)\).
The field equations then boil down to
\begin{align}
&\frac{d^2|\phi|}{du^2} + \frac{2}{u} \frac{d|\phi|}{du} + \epsilon W^2 |\phi| 
= 3s\beta e^{2\beta|\phi|^2} |\phi|H(1-u) \,,\\
&\frac{d^2W}{du^2} + \frac{2}{u} \frac{dW}{du} = \epsilon |\phi|^2 W \,.
\end{align}

Solving these equations perturbatively in $\epsilon$ to leading order, and, again,
in a small-amplitude expansion for $|\phi|$ to
sub-leading order, and 
matching at the stellar boundary, one finds that $W$ is forced to vanish 
everywhere in order to avoid a singularity at the stellar centre. As the equations
for \(|\phi|\) are identical to those of \(\phi\) in the previous model (other than the
contribution which is forced to vanish) the scalarized profile and the threshold are
unaffected. Hence, this calculation also demonstrates that scalarization will proceed in the same fashion as in the known models explored by
Damour and Esposito-Far\`{e}se.

To summarise, we have presented two models in which the mass of the photon has a different value in the interior and the vicinity of a compact star than that measured by experiments performed in a weak gravity regime. In both models a scalar field $\phi$ undergoes spontaneous scalarization, {\em i.e.}~its configuration is trivial in and around matter configurations of low compactness, whereas it becomes non-trivial once a certain threshold in compactness is crossed. The first model has a Proca-like mass term with a $\phi$ dependent effective mass, and hence scalarization can change the value of this mass from undetectably low to significantly high to give rise to new phenomenology. The second model can be thought of as a gravitational Higgs mechanism with the Higgs potential replaced by the scalar-gravity coupling. In this model the massless photon acquires a nonzero mass in the interior of a sufficiently compact star. 

We have focused on the electromagnetic field for concreteness but it should be clear that our goal was to give a proof of principle. One can straightforwardly construct similar models that would change the masses or couplings of other standard model fields in and around compact stars, {\em i.e.}~in high curvature regimes. This can have profound implications for our understanding of the microphysics and internal structure of neutron stars, as currently realistic equations of state are based on the assumption that fundamental physics remains unchanged in the star's interior. Moreover, models such as the ones we propose here could exhibit characteristic phenomenology that current or future observations could observe. This would reveal the existence of otherwise very elusive scalar fields. This issue certainly deserves closer investigation.

It has recently been shown that scalarization might also occur around black holes that are surrounded by matter \cite{Cardoso:2013fwa,Cardoso:2013opa}, as the presence of the latter makes the known no-hair theorems \cite{Chase,Bekenstein:1971hc,Hawking:1972qk,Sotiriou:2011dz,Bekenstein:1996pn,Sotiriou:2015pka,Herdeiro:2015waa} inapplicable. If significant scalarization can occur in some astrophysical black hole systems, then it would be particularly interesting to study such systems within the context of our models. 

Before closing, it is worth commenting on our choices of coupling between the scalar and the electromagnetic field. Our models are such that the coupling terms vanish at low orders in perturbation theory around the unscalarized solution and hence they are entirely absent from the equation of motion of the scalar field. This is the simplest way to be certain that scalarization is entirely unaffected and proceeds precisely as in the known models at perturbative level. However, a contribution to the scalar's equation coming from the coupling terms could well be present, so long as it does not quench the tachyonic instability that acts as a perturbative manifestation of spontaneous scalarization. Models with such behaviour are very likely to exist.

We close with a note of caution. Our arguments relied heavily on a perturbative treatment. A nonperturbative study of compact stars in our models would not only provide stronger evidence for our claims, but it would also determine the structure of compact stars and allow one to quantify the deviations from general relativity. We will consider this issue in a separate publication.

\begin{acknowledgments}
The research leading to these results has received funding from the European
Research Council under the European Union's Seventh Framework Programme
(FP7/2007-2013) / ERC grant agreement n.~306425 ``Challenging General
Relativity''.
\end{acknowledgments}




 \end{document}